\documentclass[doublecol,graphicx,times]{epl2} 
\usepackage{graphicx}
\usepackage{amsmath}
\usepackage{amsfonts}
\usepackage{amssymb}
\usepackage{amsbsy}

\def\apj{{Ast\-ro\-phys.\ J.}}

\def\grl{{Geo\-phys.\ Res.\ Lett.}}

\def\jgr{{J.\ Geo\-phys. Res.}}

\def\prl{{Phys.\ Rev.\ Lett.}}

\def\pss{{Planet.\ Space Sci.}}

\def\ssr{{Space Sci.\,Rev.}}

\def\pf{{Phys. Fluids}}
\def\npg{{Nonlin.\ Process.\ Geophys.}}

\def\pop{{Phys.\ Plasmas}}
\def\jpp{{J.\ Plasma Phys.}}
\def\prl{{Phys. Rev. Lett.}}

\title{On deformation of electron holes in phase space}
\shorttitle{Phase space hole deformation} 

\author{R. A. Treumann,\inst{1,2} C. H. Jaroschek\inst{3} \and R. Pottelette\inst{4}}
\shortauthor{R. A. Treumann, C. H. Jaroschek, and R. Pottelette}

\institute{ 
  \inst{1} Department of Geophysics, Munich University, Theresienstr. 41, D-80333 Munich, Germany\\                   
  \inst{2} Department of Physics and Astronomy, Dartmouth College, Hanover, NH 03755\\
  \inst{3} Department of Earth and Planetary Sciences, Tokyo University, Tokyo, Japan\\
  \inst{4} CETP/CNRS St. Maur des Foss\'es, Cedex, France}

\abstract{
This Letter shows that for particularly shaped background particle distributions momentum exchange between phase space holes and the distribution causes acceleration of the holes along the magnetic field. In the particular case of a non-symmetric ring distribution (ring with loss cone) this acceleration is nonuniform in phase space being weaker at larger perpendicular velocities thus causing deformation of the hole in phase space.}

\begin{document}

\maketitle
\section{Introduction}
In configuration space, phase space holes appear as localised intense electrostatic fields $E_{\| h}=-\nabla_\|\phi_h$ with broadband spectral signature parallel to the ambient magnetic field. In velocity space they form narrow regions of lacking particles of one signature kept alive for a limited time by the electrostatic field. Ion holes are local deficiencies of ions while electron holes are local deficiencies of electrons. Thus the former correspond to weak negative, the latter to positive space charges $Q_{i,e}$. In this Letter we deal with electron holes which can be excited by  beam or current instabilities parallel to the ambient magnetic field ${\bf B}$, like the two-stream instability which works for electron drifts $v_d>v_e$, larger than the electron thermal velocity $v_e$ \cite{buneman1958,buneman1959}. At lower drifts this instability is replaced by a modified version \cite{kindel1971} which is a form of the modified two-stream instability \cite{lampe1971,lampe1972,ott1972,gladd1976}. Their theory has been given by Bernstein, Greene and Kruskal \cite{bgk1957}, Schamel \cite{schamel1972,schamel1975,schamel1986}, Dupree \cite{dupree1982,dupree1983} and Turikov \cite{turikov1984}. Simulations by Newman et al. \cite{newman2001,newman2002}, Muschietti et al. \cite{muschietti1999a,muschietti1999b} and others have shown that electron holes are the natural nonlinear state of these instabilities, being Debye scale entities along  ${\bf B}$ in configuration space, and of short extension in the parallel velocity component $v_\|$ in velocity space. They contain a dilute component of trapped electrons of density $N_t$ of low energy $mv_t^2/2\leq |e\phi_h|$. In configuration space they are oblate in the direction perpendicular to the magnetic field (pancakes). Their behaviour in  $v_\perp$ has not yet been investigated in detail. It is, however, reasonable to assume that the holes are either gyro-limited, being of transverse spatial extension up to the thermal gyroradius $\Delta_{h\perp}\sim r_c=v_{e}/\omega_c$ or inertia limited $\Delta_{h\perp}\sim c/\omega_p$. Their life time is determined by the stability of the holes with respect to the generation of whistlers, trapped particle instabilities, particle trapping, heating and diffusion and the corresponding generation of dissipation (see, e.g., Newman et al. \cite{newman2002}). One might believe that these microscopic entities are of minor importance for the behaviour of the plasma. However, in collisionless plasmas they form an important dynamical source of dissipation. They heat and accelerate electrons, cause beam cooling, and are suspected to provide a substantial part of the dissipation that is needed in collisionless shocks and in reconnection. In collisionless shocks they might contribute to the emission of radiation causing the badly understood type II bursts. Some time ago we proposed  \cite{pottelette2001} that phase space holes contribute to electron cyclotron maser emission \cite{treumann2006} generating auroral kilometric radiation in the upward current source region where the holes have been identified \cite{pottelette2005}  subsequently, forming what we called `elementary radiation sources'. For this to work the holes must become deformed in phase space in order to attain a perpendicular phase space gradient $\partial F(v_\|,v_\perp)/\partial v_\perp$ on the electron distribution function, which is required by the cyclotron maser mechanism \cite{treumann2006}. A qualitative discussion of how this can be achieved has rcently been provided \cite{treumann2007}. Momentum exchange between the background electron distribution and the hole has been made responsible for deformation of the phase space shape of the hole, with the dynamics of the hole depending sensitively on the shape of the background electron distribution. In this Letter we present a  more quantitative mechanism which is developed for electron holes. However, in a similar way it should also work for ion holes in the presence of, say, ion conics, which have been found in multitude under auroral conditions. 
\section{Mechanism}
Under auroral upward current conditions the bulk distribution is  kind of a non-symmetric (downgoing) ring distribution with loss cone (due to the presence of the absorbing ionosphere) as shown in Figure 1. We assume that some appropriate instability generates an electron hole that propagates along the magnetic field ${\bf B}$ at velocity $v_h\ll V_R$ much less than the nominal electron ring velocity $V_R$. The question, in which way the hole is generated is of secondary importance for the purpose of this Letter. The qualitative discussion of \cite{treumann2007} referred to the two-stream instability \cite{buneman1958} as generator of the hole. Under the dilute plasma conditions in the upward current auroral region where $\omega_{ce}/\omega_{pe}\sim 10-25$,  the two-stream instability does not grow very fast. Its growth rate $\gamma_B\sim \omega_B \sim (m_e/16\,m_i)^\frac{1}{3}\sim 0.03\,\omega_{pe}$  is less than the global electron cyclotron maser growth rate $\gamma_{ecm}\sim (10^{-4}-10^{-3})\,\omega_{ce}$ (cf., e.g. \cite{pritchett1984,pritchett1986,pritchett2002, yoon1998}). However, hole generation is not affected by the global maser instability which has only a minor effect on the bulk electron distribution not causing a substantial energy loss for the electrons. Quasilinear flattening of the distribution which partially fills in the inner part of the ring (shown as weak background in Fig. 1) just where the hole is located in phase space is mainly caused by VLF turbulence generated under the same conditions \cite{labelle2003,treumann2006}. However, at fixed $v_\perp$ the electron drift velocity $v_d$ might not exceed the electron thermal speed $v_e$, in which case the drift cyclotron instability (MTSI) takes over. Its growth rate is of the order of the lower hybrid frequency $\gamma_{mts}\sim\omega_{lh}\sim\omega_{pi}$ which in the dilute plasma is also small. Hence, hole formation is a process comparable to or slower than the emission of radiation by the global maser instability. In the auroral kilometric radiation the latter might thus provide the background radiation level while the steep gradients produced in hole deformation \cite{treumann2007} generate the narrow intense short lived drifting emission bands that have been observed. 
\begin{figure}[t]
{\includegraphics[width=8.5cm]{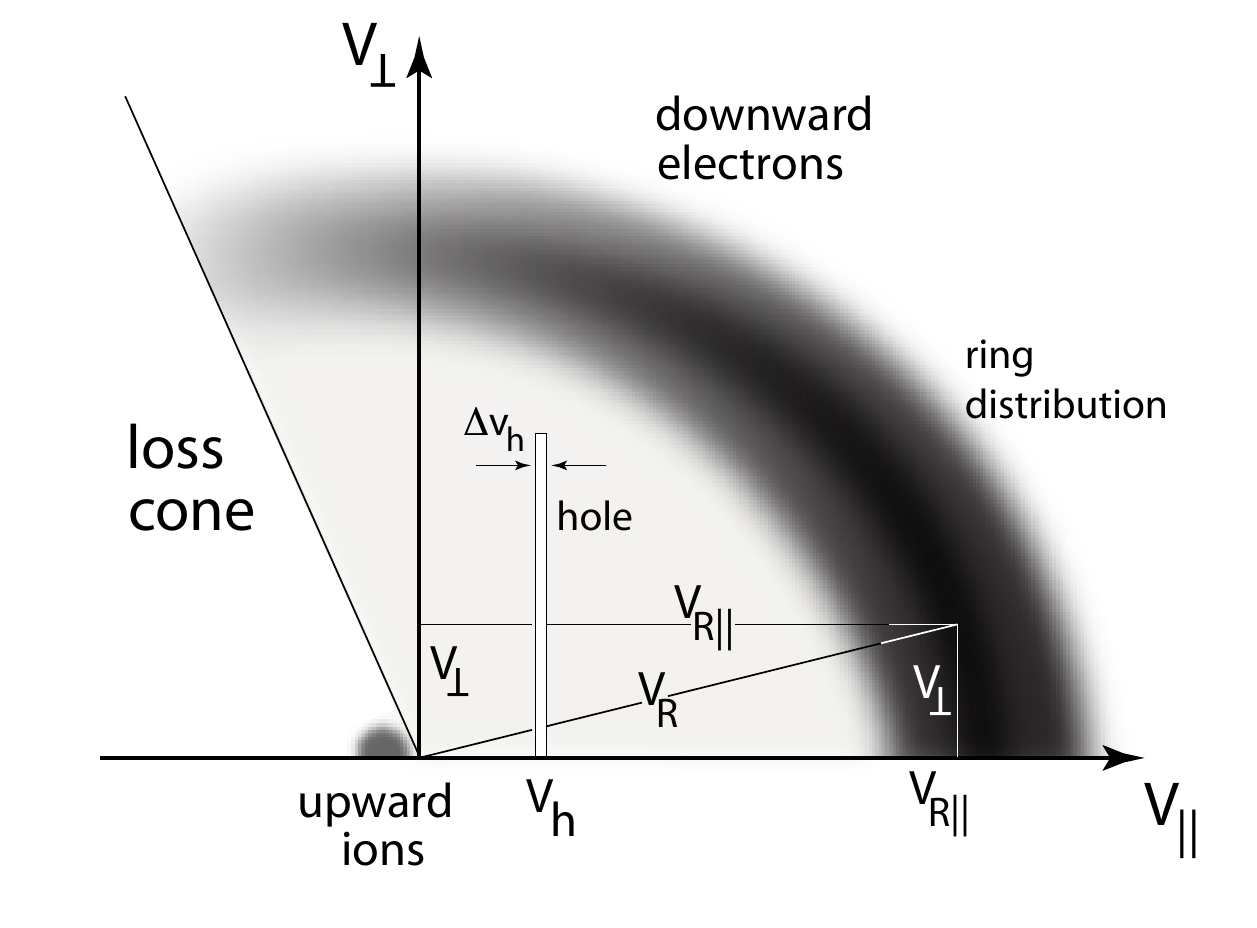}}
\caption{ Schematic of a ring distribution with loss cone. In this representation the ring is given a varying phase space density with increasing pitch angle ending at a large empty loss cone. Also shown is an electron hole at location of parallel velocity $v_h$ and for a range of perpendicular velocities $v_\perp$. The hole is assumed to be a straight line initially at constant $v_h$ being located between the hot electron ring distribution and the cold ion distribution. The latter propagates into opposite direction to the electrons at much lower speed. The hole is slow against the electrons while its velocity might be comparable to the ion speed. Also shown are the components of the nominal ring velocity $V_R$ a line of constant $v_\perp$ used in the calculations.}\label{hole3-fig1}
\end{figure}

The hole-related localised deficiency of electrons on the electron background in phase space is centred at the instantaneous velocity $v_h(t)$ of the hole (Fig. 1). In complete analogy to solid state physics it represents a localised {\it positive} charge $Q_h$ in the electron fluid that is attached to the hole. The hole moves with velocity $v_h(t)$ on the electron fluid. In the presence of an electric field $E_\|$ this charge $Q_h$ will become accelerated, with its collisionless dynamics being described by the equation of motion
\begin{equation}\label{eq-1}
\left.\frac{{\rm d}v_h(t)}{{\rm d}t}\right|_{v_\perp}=\frac{Q_h}{M_h}E_\|.
\end{equation}
Since the hole can move only parallel to the ambient magnetic field, this is a one-dimensional equation of motion that holds at every fixed perpendicular velocity $v_\perp$, as indicated on the left hand side. The interaction of the hole is purely electrostatic, and the electric field is external to the hole given by the conditions in the plasma. These are described by the presence of the downward electrons which move with respect to the ambient ions a shown in the fixed laboratory reference frame in Figure 1. Moreover, the ion and electron beams are distributed about homogeneously over an area that is large against the extension of the hole in any direction. This implies that the electric field can be expressed through the current flowing in the plasma as 
\begin{equation}\label{eq-2}
\frac{\partial E_\|}{\partial t} = \left.\frac{eN}{\epsilon_0}v_d\right|_{v_\perp}, \qquad {\rm where} \qquad v_d=\langle v_{e\|}\rangle-\langle v_{i\|}\rangle.
\end{equation}
Here $N$ is the plasma density. The drift speed is the difference between the average bulk speeds of the electrons and ions. In order to find an equation for the parallel velocity of the hole, we take the time derivative of Eq. (\ref{eq-1}), remembering that the hole charge $Q_h(v_h)$ itself is a function of the hole speed. Defining $q=Q_h/e$ and $m=M_h/m_e$ this yields
\begin{equation}\label{eq-3}
\frac{{\rm d}^2v_h}{{\rm d}t^2}-\frac{{\rm d}\ln q (v_h)}{{\rm d}v_h}\left(\frac{{\rm d}v_h}{{\rm d}t}\right)^{\!\!2}=\omega_{pe}^2v_d\left[\frac{q(v_h)}{m}\right].
\end{equation}
In deriving this expression we used Eqs. (\ref{eq-1}-\ref{eq-2}) in order to eliminate the electric field, and we have suppressed the index $v_\perp$ understanding that Eq. (\ref{eq-3}) holds for every constant perpendicular velocity. To be able to proceed we must determine the charge and mass of the hole. The total mass of the hole is given by $M_h=m_eN_tV_h$, with $N_t$ the number density of electrons that are trapped in the spatial volume $V_h$ of the hole. This number is assumed to be constant during the evolution of the hole, a simplification which neglects any possible exchange of electrons between the hole and its environment which determines its life time. Hence over the entire evolution of the hole $m=$\,const does not change. This is not so for the charge of the hole $Q_h=e(\left.N\right|_{v_h}-N_t)V_h$. Any acceleration of the hole in real space corresponds to a displacement in velocity space. Hence  $Q_h$ depends on the fraction in ambient number density of ambient electrons of different velocity seen by the hole when it moves in velocity space. We need to know only the ratio 
\begin{equation}\label{eq-4}
q/m=(\left.N\right|_{v_h}-N_t)/N_t,
\end{equation}
and, hence, the volume $V_h$ of the hole cancels out leaving us with the problem to calculate $\left.N\right|_{v_h}$. In order to do this we assume, for simplicity, that the background ring distribution is a rotated Maxwellian
\begin{equation}\label{eq-5}
F(v_\|,v_\perp)=\frac{N}{\pi^{\frac{3}{2}}v_e^3}\exp\left\{-\frac{({\bf v}-{\bf V}_R)^2}{2v_e^2}\right\}.
\end{equation}
Here $V_R=$ const is a constant radius in velocity space, and ${\bf v}=(v_\|,v_\perp)$ is the velocity vector. (In global electron-cyclotron maser theory and simulations  as, e.g., in \cite{pritchett2002}, the thermal width of the distribution is not important and a $\delta$-ring distribution is used.) At constant $v_\perp$ the variation is in the parallel component $(v_\|-V_{R\|})^2$ in the argument of the exponential. The fractional number density of the hole is obtained by integration just over the velocity volume of the hole. Since this is very small we assume that the hole is of rectangular shape of width $\Delta v_h=\sqrt{2\phi_h/m_e}$ in parallel velocity. Then
\begin{equation}\label{eq-6}
\left.N\right|_{v_h}=2\pi\Delta v_h \left.F\right|_{v_h}v_\perp{\rm d}v_\perp.
\end{equation}
The indication $v_h$ on $F$ means that in $F$ the parallel velocity $v_\|\to v_h$ is to be replaced by the hole speed. It is then easy to show that the coefficient of the second term on the left in Eq.\,(\ref{eq-3}) is ${\rm d}\ln q/{\rm d}v_h=-(v_h-V_{R\|})/v_e^2$, which is linear in $v_h$. This fact would enable us to rewrite Eq.\,(\ref{eq-3}) in terms of the variable $x=v_h-V_{R\|}$. However, it is more convenient to introduce dimensionless variables $\tau=\omega_{pe}t, u=v_h/v_e, U_R=V_{R\|}/v_e, U_d= v_d/v_e, n=N/N_t$ and rewrite Eq.\,(\ref{eq-3}) as
\begin{equation}\label{eq-7}
u''-(U_R-u)(u')^2=(U_d-u)\left\{nC\,{\rm e}^{-\frac{1}{2}(U_R-u)^2}-1\right\},
\end{equation}
where $'={\rm d}/{\rm d}t$, and $C$ for fixed $v_\perp$ is a constant factor that is determined from the background distribution $F$. The right-hand side of this equation is linearly proportional to the average parallel beam velocity $U_d$ which is the driving force. $U_d=0$ corresponds to a symmetric ring distribution with no loss-cone. The rigth-hand side vanishes for $U_d=u$, in which case the equation is solved trivially (expressed in decaying error functions in $u$), and it can be shown that an initially finite velocity hole will come to rest. Generally one may suspect that asymptotically $u\to U_d$. Hence, for a non-symmetric ring distribution that, for instance, involves a loss cone, the section of the hole corresponding to $v_\perp=$ const is accelerated parallel to ${\bf B}$ until settling near $U_d$. This causes deformation of the hole in phase space and is described by the solutions of Eq.\,(\ref{eq-7}) for the most interesting case $U_d\gg u$.

The hole is a slowly moving entity of velocity $u\ll U_R$. Observation of the displacement of the fine structure in the auroral kilometric radiation suggests that holes move at velocity $v_h\lesssim 100$ km/s \cite{pottelette2001} compared to electron beam velocities of $V_{R\|}\sim 10^4$ km/s. Hence, in the coefficient of the second term on the left in Eq.\,(\ref{eq-7}) $u$ can be neglected. Moreover, the expression in front of $U_d-u$ on the right-hand side becomes a constant $A\equiv nC\exp(-U_R^2/2)$. If we assume that $U_d\gg u$, which holds for a non-symmetric electron distribution like that shown in fig.1, the solution of Eq.\,(\ref{eq-7}) can be found by multiplication with $u'$ and integrating once with respect to ${\rm d}\tau$. This yields the following expression for
\begin{equation}\label{eq-8}
u'(\tau)=\sqrt{AU_d/U_R}\tan\left[\sqrt{AU_RU_d}(\tau_f-\tau)\right],
\end{equation}
with $\tau=\tau_f$ the final time when the hole comes to rest and $u'=0$. The restriction on the argument of the tangent function implies the following restriction on $\tau_f$:
\begin{equation}\label{eq-9}
\tau_f\lesssim\pi\left(2\sqrt{AU_RU_d}\right)^{\!\!-1}.
\end{equation}
The larger the electron drift velocity $U_d$, the faster the hole tends to reach its final velocity. In other words, the faster the electron beam moves, the less time it takes for the deformation of the hole in phase space. This is what has been expected from the very beginning. Eq.\,(\ref{eq-8}) also shows that the acceleration of the hole is a monotonically decreasing function of time $\tau$ until the hole arrives at its final state. The velocity $u(\tau)$ of the hole is obtained by integrating Eq.\,(\ref{eq-8}) with respect to time. In the initial state for times $\tau$ short against $\tau_f$ a solution is found by expanding the integrand. Restricting to the first two terms yields
\begin{equation}\label{eq-10}
u(\tilde\tau)\sim \frac{\tilde\tau}{U_R}\left\{ \tan{\tilde\tau_f}-\frac{1}{2}\frac{\tilde\tau}{\cos^2\tilde\tau_f}\right\},
\end{equation}
where $\tilde\tau=\tau\sqrt{AU_RU_d}$. Initially the hole velocity increases linearly with time, and the acceleration is proportional to $\sqrt{U_d}$, the root of the electron drift velocity and vanishes for $U_d\to 0$ in the case of a symmetric electron ring confirming the former conclusion on this case. 

In the more general case when $u$ is of the same order as $U_R$, Eq.\,(\ref{eq-7}) cannot be solved analytically. Since this case is of lesser interest and would exceed the limits of a Letter, we leave its investigation for a later publication (Jaroschek et al., to be submitted). Instead we proceed to investigate the variation of the hole velocity with the shape of the electron distribution. This is closely related to the variation with $v_\perp$ and the drift velocity $U_d$. To this end we study the effect of $A=nC\exp(-U_R^2/2)-1$ respectively $C$. The constant $C$ is a differential in $v_\perp$
\begin{equation}
C=\frac{2\Delta v_h}{\sqrt{\pi}v_e}\exp\left[-\frac{(v_\perp-V_{R\perp})^2}{2v^2_e}\right]\frac{v_\perp{\rm d}v_\perp}{v_e^2}.
\end{equation}
At a phase space section parallel to $v_\|$ all particles in the distribution have the same $v_\perp=V_{R\perp}$, and the exponential factor in $C$ is unity. We may define the electron drift speed and electron temperature at constant $v_\perp$ as the first and second moments of the electron distribution function at constant $v_\perp$, respectively.  The local drift velocity entering into the expressions for the evolution of $u(\tau)$ is the parallel component of the first moment of the electron distribution function at constant $v_\perp$. For a half ring distribution  with $V_{R\|}>0$ one has
\begin{equation}
\left.U_d\right|_{v_\perp}^+=2^\frac{3}{2}\left(\frac{V_R^2}{v_e^2}-\frac{v_\perp^2}{v_e^2}\right)^{\!\!\frac{1}{2}}\frac{v_\perp{\rm d}v_\perp}{v_e^2},
\end{equation}
an expression that decreases with increasing $v_\perp$. The same expression holds for the part of the ring in Fig. 1 that has no expression at negative velocities in the loss cone on the left of Fig. 1. The decrease in $U_d$ with $v_\perp$ is crucial for the evolution of the parallel velocity $u(\tau)$ of the hole in phase space in the presence of a half ring. Since the hole velocity and acceleration of the hole in parallel direction are proportional to $U_d$ the decrease implies that the largest variation in $u$ is for small $v_\perp$. The velocity space deformation decreases with increasing perpendicular velocity $v_\perp$. Hence, the hole attains the largest velocity at small $v_\perp$ thus becoming bent in velocity space, as was suggested earlier \cite{treumann2007} from qualitative considerations. 

If the loss cone is not empty, the effective drift velocity $ U_d=U_d^+ -U_d^-$ the hole experiences is reduced by the contribution of the ring at negative velocities
\begin{equation}
\left.U_d\right|_{v_\perp}^-= 2^\frac{3}{2}\alpha\left(\frac{V_R^2}{v_e^2}-\frac{v_\perp^2}{v_e^2}\right)^{\!\!\frac{1}{2}}\frac{v_\perp{\rm d}v_\perp}{v_e^2},
\end{equation}
where $\alpha$ is the fractional density  of the half-ring distribution with $V_{R\|}<0$, while the inclusion of the upward ion distribution at low $v_\perp$ would increase the effective drift speed for the small values of $v_\perp$ to which the ions extend in velocity space (see Fig. 1). 

The case $U_d<u$ is not of vital interest. In this case $u\to 0$, and the hole will be about at rest. Such cases may refer to standing narrow band emissions in the auroral kilometric radiation with the loss-cone about filled.

\section{Discussion}
The above considerations show that an electron phase space hole experiences a particular dynamics in phase space that depends on the shape of the distribution function. This Letter was restricted to the mere investigation of the possibility of deformation of an initially straight (in velocity space) phase space hole (BGK mode) by momentum exchange with the ambient electron distribution. We argued that momentum exchange causes differential acceleration of the hole different for different $v_\perp$. 

The evolution of phase space holes has so far been considered in theory and simulations only for Maxwellian distributions in which case their phase space dynamics is simple (see, e.g., \cite{singh2000,newman2002}). Observations in space, for instance under conditions in the auroral kilometric radiation source and also in collisionless shocks, suggest that phase space holes evolve when the phase space particle distributions deviate strongly from Maxwellian shape \cite{delory1998,ergun2000,ergun2001,ergun2002}. In particular, the combination of an electrostatic field along the magnetic field and a magnetic mirror geometry transform a beam distribution into a ring distribution \cite{chiu1978} with loss cone (colloquially called a horseshoe). Such distributions are the rule in the aurora and are also expected in super-critical quasi-perpendicular collisionless shocks \cite{treumann2008}. In their presence a hole should undergo deformation in phase space of the kind described in this Letter. Holes have also been predicted in relation to reconnection in collisionless current sheets \cite{drake2003} where their signatures might have been observed {\it in situ} in space \cite{vaivads2004,sundkvist2007}. 

The phase space deformation might not be of overwhelming importance for the dynamics of the plasma even though its importance for dissipation has not yet been investigated. However, as suggested in \cite{treumann2007} it should be of crucial importance for involving electron holes into the emission of electromagnetic radiation by the electron-cyclotron maser mechanism acting in the auroral regions of planetary magnetospheres and in a variety of astrophysical objects (cf., e.g., \cite{begelman2005}). This requires the generation of sharp positive phase space gradients $\partial F(v_\||,v_\perp)/\partial v_\perp >0$ on the electron distribution function in the perpendicular velocity component. Phase space holes do not originally possess such gradients and therefore should not become involved into generation of radiation other than contributing to electron heating and acceleration of electron beams which on their own might secondarily generate plasma waves away from the hole and produce second harmonic plasma radiation with frequency $2\omega_{pe}$. This mechanism is viable, for instance, at collisionless shocks and might be responsible for the so-called backbone radiation in type II radio bursts observed in the solar corona and interplanetary space. 

However, for electron holes to become directly involved into the electron cyclotron maser mechanism, generation of perpendicular gradients is absolutely necessary. This is provided by the phase space deformation mechanism proposed in this Letter, which bends the hole in phase space and transforms its natural parallel phase space gradient $\partial F(v_\|,v_\perp)/\partial v_\|\neq 0$ into a perpendicular gradient. Since $\partial F(v_\|,v_\perp)/\partial v_\|\neq 0$ is steep and increases when the hole is accelerated into the bulk of the distribution \cite{treumann2007}, the new $\partial F(v_\||,v_\perp)/\partial v_\perp$ is also very steep and will readily contribute to maser emission. Its frequency maps the local electron cyclotron frequency or its lower harmonics. We are not going here into the details of the radiation process as this has been described in the literature for the global electron distribution (see e.g. \cite{pritchett1984,louarn1990,louarn1996}). Since the hole possesses two boundaries, one of them causes a positive, the other a negative phase space gradient. It has been argued \cite{treumann2007} that, theoretically, these gradients correspond to emission and absorption separated by the whole width. Current instrumental resolution and possibly natural line broadenings do not allow for a discrimination of these emissions and absorptions, however. Whistler saucer emissions from holes might be another indication of evolution of positive perpendicular phase space gradients.

The emission frequency changes when the hole is displaced in space along the magnetic field. Because of the steepness of the gradient, the emission is very narrow band. From its spectral displacement the electron hole speed can be determined while from its spectral width properties of the hole can be inferred \cite{treumann2007}. In the auroral region the hole speed is found to be of the order of $\lesssim 100 $ km/s, which is comparable to the ion-acoustic speed being sufficiently far below the electron thermal velocity to justify the model of a slow hole. However, the measured spectral speeds reflect the bulk velocity of the hole. This is determined by integrating over the hole distribution along the bent phase space shape of the hole. It can be much less than the differential speeds of its fastest moving small $v_\perp$ phase space sections. 

We also note that ion holes behave in a similar way experiencing bending in phase space if the ion distribution differs from a Maxwellian. This is the case in regions where ion conics are generated in the presence of field aligned electric potentials and mirror geometries for upward going ions. The evolving steep perpendicular velocity space gradient may excite ion cyclotron waves from ion holes. 

\acknowledgments This research is part of a Visiting Scientist Programme at ISSI, Bern. CHJ acknowledges a JSPS Fellowship of the Japanese Society for the Promotion of Science. CHJ and RT thank M. Hoshino for hospitality, support and discussions. This research has also benefitted from a Gay-Lussac-Humboldt award of the French Government.

\end{document}